\documentclass[sort&compress,AMA,STIX1COL]{WileyNJD-v2}
\articletype{Research Article}%
\received{26 April 2016}
\revised{6 June 2016}
\accepted{6 June 2016}
\usepackage{comment}
\usepackage{algorithm}
\usepackage{multirow}
\usepackage{array}
\usepackage{float}
\usepackage{url}
\usepackage{xcolor}

\begin{document}
\title{Assurance methods for designing a clinical trial with a delayed treatment effect}
\author[Salsbury]{James A. Salsbury*}
\author[Oakley]{Jeremy E. Oakley}
\author[Julious]{Steven A. Julious}
\author[Hampson]{Lisa V. Hampson}
\authormark{SALSBURY \textsc{et al}}
\address[Oakley, Salsbury]{\orgdiv{The School of Mathematics and Statistics}, \orgname{The University of Sheffield},
\orgaddress{\country{U.K.}}}
\address[Julious]{\orgdiv{The School of Health and Related Research}, \orgname{The University of Sheffield},
\orgaddress{\country{U.K.}}}
\address[Hampson]{\orgdiv{Advanced Methodology and Data Science}, \orgname{Novartis Pharma AG, Basel},
\orgaddress{\country{Switzerland}}}
\corres{*James Salsbury, The School of Mathematics and Statistics, The Hicks Building, Broomhall, Sheffield, S3 7RH
\email{jsalsbury1@sheffield.ac.uk}}
\abstract{An assurance calculation is a Bayesian alternative to a power calculation. One may be performed to aid the planning of a clinical trial, specifically setting the sample size or to support decisions about whether or not to perform a study. Immuno-oncology is a rapidly evolving area in the development of anticancer drugs. A common phenomenon that arises in trials of such drugs is one of delayed treatment effects, that is, there is a delay in the separation of the survival curves. To calculate assurance for a trial in which a delayed treatment effect is likely to be present, uncertainty about key parameters needs to be considered. If uncertainty is not considered, the number of patients recruited may not be enough to ensure we have adequate statistical power to detect a clinically relevant treatment effect and the risk of an unsuccessful trial is increased. We present a new elicitation technique for when a delayed treatment effect is likely and show how to compute assurance using these elicited prior distributions. We provide an example to illustrate how this can be used in practice and develop open-source software to implement our methods. Our methodology has the potential to improve the success rate and efficiency of Phase III trials in immuno-oncology and for other treatments where a delayed treatment effect is expected to occur.}
\jnlcitation{\cname{%
\author{James A. Salsbury}
\author{Jeremy E. Oakley},
\author{Steven A. Julious}, and
\author{Lisa V. Hampson}} (\cyear{<year>}),
\ctitle{<journal title>}, \cjournal{<journal name>} <year> <vol> Page <xxx>-<xxx>}
\maketitle

\section{Introduction}
     
Assurance calculations are growing in popularity as an aid for the design of clinical trials. An assurance calculation is a Bayesian alternative to a power calculation: instead of assuming parameters (eg related to treatment effects) take particular values, we elicit prior distributions for them, enabling us to derive a probability of a successful trial outcome, accounting for uncertainty about treatment effects. The concept of an assurance calculation was first considered by Spiegelhalter and Freedman,\cite{Spiegelhalter1986} then developed by O’Hagan et al,\cite{OHagan2005} who coined the term `assurance'. Note that the assurance method has had other terms accredited to it, such as average power, expected power and predictive power.\cite{Grieve2022} 

To calculate assurance, we sample from the elicited prior distributions for the unknown parameters and then simulate clinical trials using these sampled values. The prior predictive probability that the trial will be `successful' is the proportion of simulated trials that meet our stated success criteria. These success criteria are not fixed by the assurance method; instead, they are set independently by the sponsor and can be any criteria that the sponsor wishes to consider (eg, that the observed treatment effect will be positive; or statistically significant; or exceed a clinically relevant threshold). More recently, assurance has been used to calculate the probability of obtaining regulatory approval with clinically relevant effects on key endpoints after Phase IIb.\cite{Hampson2022a, Hampson2022b}. Note that the method for the trial data analysis is also specified independently of the assurance method; the same analysis method would be assumed as that used in the power calculation.  

Assurance provides a more realistic assessment of the probability a trial will give rise to a successful outcome compared to a conventional power calculation. The high failure rates of clinical trials are well-documented\cite{Gotte2015} and there are several examples where promising results from early phase trials have not been replicated in subsequent Phase III trials. \cite{FDA2017} Assurance calculations for Phase III trials should capture the strength of the available evidence after mitigating the selection bias often inherent in early phase data when a necessary condition for progress is positive Phase Ib or Phase II results. They should also account for any limitations in the available data in light of planned shifts in the patient population, outcome or treatment strategies between phases. 

Accurate and reliable evaluations of risk can be used to optimize trial design and analysis plans. For example, assurance can be used to support decisions regarding study sample size, and quantitatively measure how effective various trial setups are at reducing risk, such as the timing and number of planned interim analyses.\cite{Dallow2018, Crisp2018} Furthermore, assurance evaluations can also enable better informed decisions on whether or not to conduct a study. Of course sponsors, such as pharmaceutical companies and public funding bodies, may choose to fund a Phase III clinical trial (or indeed a program of Phase III clinical trials) regardless of whether it has a low assurance if the corresponding expected net present value (eNPV) is sufficiently high, thus targeting resources towards research programs with the greatest expected impact for patients. 

Immuno-oncology (IO) is a rapidly evolving area in the development of anticancer drugs. In trials of IO therapies, time-varying treatment effects that deviate from the proportional hazards (PH) assumption have been observed on time-to-event endpoints such as progression-free survival (PFS) and overall survival (OS). See for example, CheckMate 017.\cite{Brahmer2015} In a systematic review of 63 confirmatory randomized controlled trials (RCTs) of anti-programmed cell death protein-1 and anti-programmed death/ligand 1 therapies, \cite{Mukhopadhyay2022} 15 studies were identified with suspected nonproportional hazards due to reasons including crossing of the OS survival curves\cite{Rizvi2020, Herbst2020} or a lag before the PFS survival curves separated.\cite{Shitara2020} In what follows, we focus on the latter scenario and refer to this as a delayed treatment effect (DTE). 

There are several challenges associated with the design and analysis of trials with nonproportional hazards. Firstly, the primary estimand should be defined with a clinically interpretable measure used to summarize the benefit of the test treatment versus control,\cite{EstimandFW2019} and an unbiased estimator should be selected to target it. Secondly, the test of the null hypothesis of no benefit of treatment versus control should be carefully selected acknowledging the impact of potential deviations from PH on the attained power of commonly applied procedures, such as the log-rank test.\cite{Jimenez2019} Where we suspect (but are not certain) that there will be a delay in the treatment effect, and furthermore are uncertain about the length of the delay if there is one, the target event number and corresponding sample size needs to be carefully chosen to provide confidence the trial will be able to meet its objectives in light of these uncertainties.

As IO trials are becoming more common, so are trials in which a DTE is observed. However, to the best of our knowledge, there has been no published work on eliciting prior distributions and calculating assurance for when a DTE is likely to be present in a clinical trial with time-to-event endpoints. In this article, we propose a method for how to elicit the relevant parameters for this trial and how to perform an assurance calculation.

In Section \ref{sec:ass}, we briefly discuss the assurance method and how it is used in practice. In Section \ref{sec:methods}, we define DTEs, present an elicitation method and signpost to the open-source software we have developed for use in this situation. In Section \ref{sec:example}, we illustrate how our method can be used to calculate assurance. In Section \ref{sec:simpgamma}, we investigate the robustness of our parameterisation and lastly we conclude with a brief summary in Section \ref{sec:summary}.

\section{Assurance}\label{sec:ass}
Suppose that an RCT is to be conducted to compare an experimental treatment with a control; we assume that this is the current standard of care, but could also be a placebo. We want to test the null hypothesis $H_0$ that the treatment effect $\theta=0$ versus the alternative hypothesis $H_1$ that $\theta \neq 0$. For a power calculation, the sample size is chosen to solve
\begin{equation}\label{eq:power}
    P(\text{Reject } H_0|\theta = \theta_A) = \pi^{*},
\end{equation}
for some desired probability $\pi^{*}$ (usually 80\% or 90\%) and treatment effect $\theta_A$, typically chosen to represent a plausible and clinically relevant effect.

The power of the test of $H_0$ at $\theta_A$ is the probability of rejecting $H_0$ if $\theta$ is as large as $\theta_A$. However, since the $\theta$ may differ from $\theta_A$, the attained power of the test may deviate from the target $\pi^{\star}$. Assurance is the unconditional probability that the trial will end with the desired outcome:
\begin{equation}\label{eq:psuccess}
    P(\text{`Successful trial'}) = \int P(\text{`Successful trial'}|\theta)f(\theta)\text{d}\theta,
\end{equation}
where $f(\theta)$ is the prior distribution for $\theta$. If a successful trial simply corresponds to rejecting $H_0$, Equation \ref{eq:psuccess} is the \textit{expected power}, interpreting $\theta_A$ in Equation \ref{eq:power} as the true value of the treatment effect, rather than some minimum clinically relevant difference. 

If the desired outcome is to reject $H_0$ with data which favours the experimental treatment, then the event `successful trial' may be defined as `Reject $H_0$ with $\hat{\theta} > 0$'. When calculating assurance, a key question is how to define the prior distribution $f(\theta)$ for the unknown treatment effect. One approach would be to take $f(\theta)$ as the posterior distribution for $\theta$ resulting from using clinical data from an early Phase II trial to update a weakly informative prior distribution. However, this approach may fail to incorporate other sources of relevant information, and may become challenging if there are differences between the treatment effect studied in Phase II and the quantity of interest in the future trial. Alternatively, the prior distribution(s) for the parameters of interest could be elicited from a group of experts, in light of the Phase IIb trial data and any other information that is deemed relevant – data from drugs with a similar mechanism of action, knowledge about the disease area etc. For a detailed discussion about the method of eliciting parameters in these contexts, see Dallow et al.\cite{Dallow2018}

The reason expert elicitation is useful in these circumstances is to bridge the gap between data from the completed Phase IIb trial and the quantities of interest in the planned Phase III trial. For example, the future trial may consider different endpoints, the patient population may change, or a different dose/dosing regime may be proposed.\cite{Holzhauer2022} Also, when working in a rare disease setting, there may be limited data available. In this context, expert elicitation is useful as it allows the study team to combine heterogeneous sources of information (RCTs, case series, observational data) when a formal mathematical synthesis of these data would be very complex.\cite{Hampson2015}

In the context of assurance methods, O'Hagan et al\cite{OHagan2005} considered eliciting beliefs for clinical trials with Normally distributed and dichotomous endpoints. Gasparini et al\cite{Gasparini2013} also considered Normally distributed endpoints, and Alhussain and Oakley\cite{Alhussain2020} considered eliciting uncertainty about the variance of Normally distributed endpoints. For time-to-event outcomes,  Spiegelhalter et al \cite{Spiegelhalter2004} considered stipulating a Normal prior distribution for the log hazard ratio under a PH assumption, as did Hiance et al.\cite{Hiance2009} Ren and Oakley\cite{Ren2014} considered expert elicitation for both parametric and non-parametric models. Azzolina et al\cite{Azzolina2021} produced a comprehensive literature review of assurance methods that use expert elicitation (both theoretical and applied).

\section{Methods}\label{sec:methods}
\subsection{Delayed treatment effects}

\begin{figure}
\centering
\includegraphics[width=0.5\textwidth]{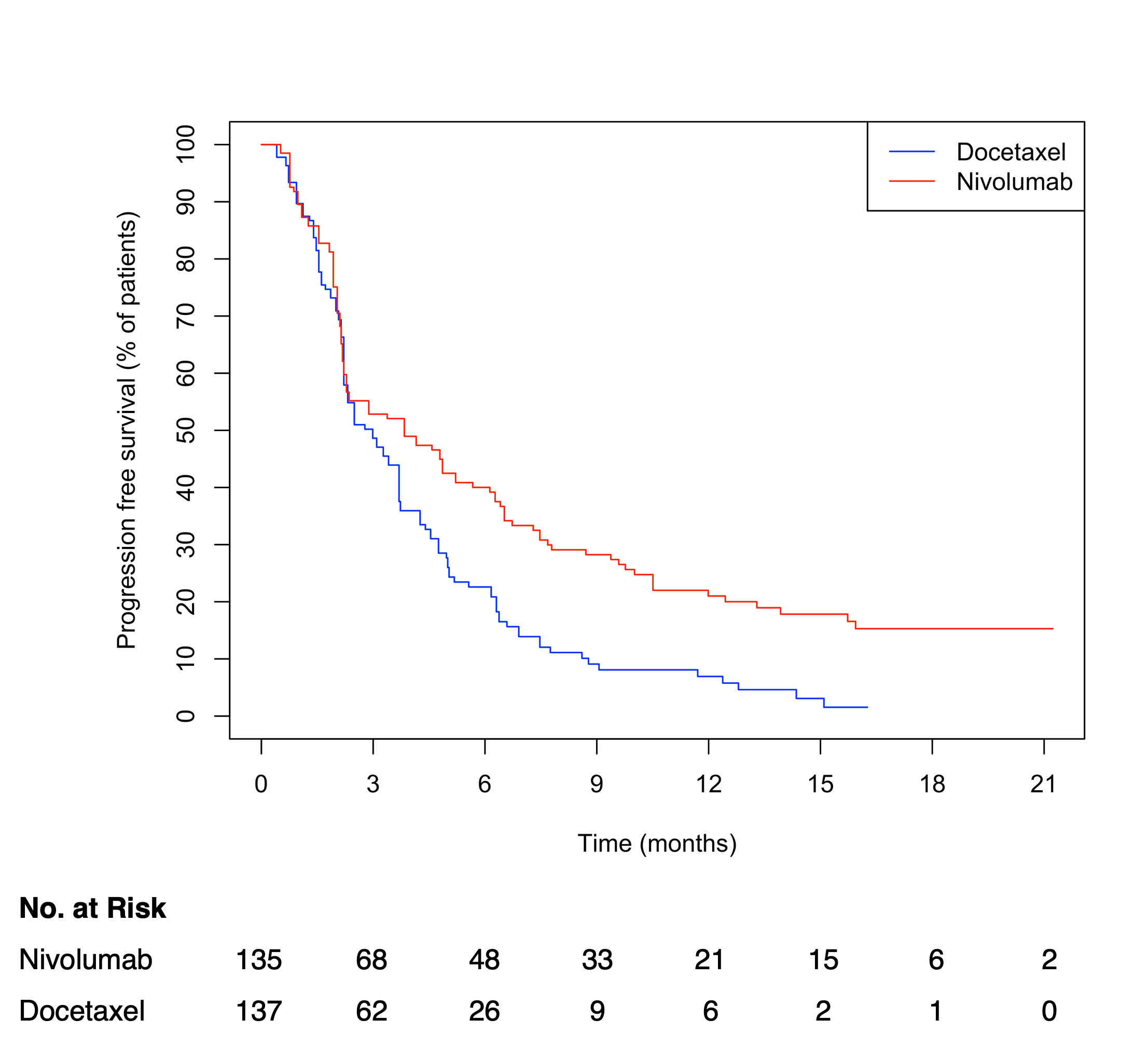}
\caption{Kaplan-Meier plot of a Phase III trial, Checkmate 017,\cite{Brahmer2015} in which DTEs are present. The control and experimental treatment curves follow the same trajectory for approximately 3 months, after which they separate.}
\label{fig:KM}
\end{figure}

Figure \ref{fig:KM} shows a Kaplan-Meier plot from a Phase III trial, CheckMate 017,\cite{Brahmer2015} in which a DTE was observed. The trial enrolled patients with advanced squamous-cell non-small-cell lung cancer (NSCLC) and compared the current standard of care, docetaxel, against an experimental treatment, nivolumab. The plot is based on the reconstructed individual patient data\cite{Liu2021} derived from published Kaplan-Meier survival curves. We see that both the control and experimental treatment curves follow the same trajectory for some time (approximately 3 months), after which they separate.

In a survival trial, suppose we have two groups: the control group and the experimental treatment group. We denote the hazard function for the control group as $h_c(t)$ and the hazard function for the experimental treatment group as $h_e(t)$. In a typical survival trial, $H_1$ assumes that the hazard function for the experimental treatment group is less than or equal to the hazard function for the control group at all time points, that is $H_1$: $h_e(t)\leq h_c(t)$, $\forall t$. This suggests that patients in the experimental treatment arm immediately benefit from the intervention compared to those in the control arm. 

In a trial in which a DTE is thought likely to occur, we make a different assumption. We assume that the hazard function for the experimental treatment group is the same as that of the control group until a certain time $T$, which represents the delay in the experimental treatment taking effect. After time $T$, we assume that the experimental treatment group starts experiencing some benefit relative to the control group:
\begin{equation}\label{eq:HF}
h_e(t) = 
\begin{cases}
h_c(t), & t \leq T \\
h^{*}_e(t) , & t > T   
\end{cases},
\end{equation}
where $h_e^*(t)\leq h_c(t)$ describes the benefit of the experimental treatment relative to control.\cite{Fine2007}

In survival trials, the PH assumption is often made. It is used in Cox regression and the log-rank test, which is a standard statistical test in survival trials, is most powerful under this assumption. However, when DTEs are present, this assumption is violated because the hazard ratio becomes time-dependent. This poses challenges for the design and analysis of trials with DTEs.

Various researchers have proposed methodologies to address trials with a DTE.\cite{Zhang2009, Xu2017, Lakatos1988, Zucker1990, Luo1994, Ristl2021,Fine2007,Sit2016, Mukhopadhyay2020,Chen2013,Li2019} However, most of these discussions focus on regaining statistical power lost due to the delay by using alternative analysis methods, such as weighted log-rank tests or the difference in restricted mean survival times (RMST).\cite{Royston2013} These methods aim to account for the time-dependent hazard ratio without assuming PH or the specific shape of the underlying survival curves.

\subsection{Assurance for delayed treatment effects}
We propose an elicitation technique and parameterisation to calculate assurance in these circumstances. By doing so, we are able to capture experts' uncertainty about the relevant parameters and provide a more realistic judgement of the probability of success of the proposed trial.

We suppose that the survival times in the control group follow a Weibull distribution with hazard function 
\begin{equation}\label{eq:HFc}
h_c(t) = \gamma_c\lambda_c^{\gamma_c} t^{\gamma_c-1} 
\end{equation}
and corresponding survival function
\begin{equation}\label{eq:control}
S_c(t) = \text{exp}\{-(\lambda_ct)^{\gamma_c}\}. 
\end{equation}

We assume survival times in the experimental treatment group, after a delay of length $T$, also follow a Weibull distribution with different parameters to the control. This induces the hazard function
\begin{equation}\label{eq:HFt}
h_e^*(t) = \gamma_e\lambda_e^{\gamma_e} t^{\gamma_e-1} 
\end{equation}
and corresponding survival function
\begin{equation}\label{eq:survtreatment}
S_e^*(t) = \text{exp}\{-(\lambda_cT)^{\gamma_c} - \lambda_e^{\gamma_e}(t^{\gamma_e}-T^{\gamma_e})\}. 
\end{equation}

\noindent Prior to time $T$, we assume the experimental treatment group has the same survival function as the control (Equation \ref{eq:control}). Thus, the survival function for the experimental treatment group is 

\begin{equation}\label{eq:treatment}
S_e(t) = 
\begin{cases}
\text{exp}\{-(\lambda_ct)^{\gamma_c}\}, & t\leq T\\
\text{exp}\{-(\lambda_cT)^{\gamma_c} - \lambda_e^{\gamma_e}(t^{\gamma_e}-T^{\gamma_e})\}, & t > T
\end{cases}.
\end{equation}

\noindent The hazard ratio of the two groups (derived from Equations \ref{eq:HFc} and \ref{eq:HFt}) is
\begin{equation}\label{eq:HR}
\text{HR}(t) = 
\begin{cases}
1, & t \leq T \\
\frac{\gamma_e\lambda_e^{\gamma_e} t^{\gamma_e-1} }{\gamma_c\lambda_c^{\gamma_c} t^{\gamma_c-1} }, & t > T
\end{cases}.
\end{equation}

\subsection{Constructing the prior distributions}\label{sec:conspriors}

From Equations \ref{eq:control} and \ref{eq:treatment}, we see that there are five unknowns: $T$, $\lambda_c$, $\gamma_c$, $\lambda_e$ and $\gamma_e$. To calculate assurance, prior distributions are required for these parameters. In the following sections we propose a method for eliciting these priors, including the questions to ask. 

\subsubsection{\texorpdfstring{Prior(s) for $\lambda_c$ and $\gamma_c$}{Prior(s) for lambda\_c, gamma\_c}}\label{sec:controlpriors}

We first elicit judgements on the two parameters for the survival times in the control group; $\lambda_c$, also known as the scale parameter, and $\gamma_c$, the shape parameter. We assume that there exists some historical data on the control so that we can derive
\begin{equation*}
    \pi(\lambda_c, \gamma_c|\boldsymbol{x}_{\text{hist}}),
\end{equation*}
where $\boldsymbol{x}_\text{hist}$ is the historical data for the control group intervention. Schmidli et al\cite{Schmidli2014} consider using a meta-analytic-predictive (MAP) prior for control group parameters. Bertsche et al\cite{Bertsche2019} extend this method to specifically consider time-to-event data. Alternatively, to include expert elicitation at this stage, see Ren and Oakley,\cite{Ren2014} who consider eliciting beliefs when survival times are assumed to follow a Weibull distribution. 

\subsubsection{Prior for \texorpdfstring{$T$}{T}}\label{sec:priorT}

\begin{figure}[h]
    \centering
    \includegraphics[width=0.65\textwidth]{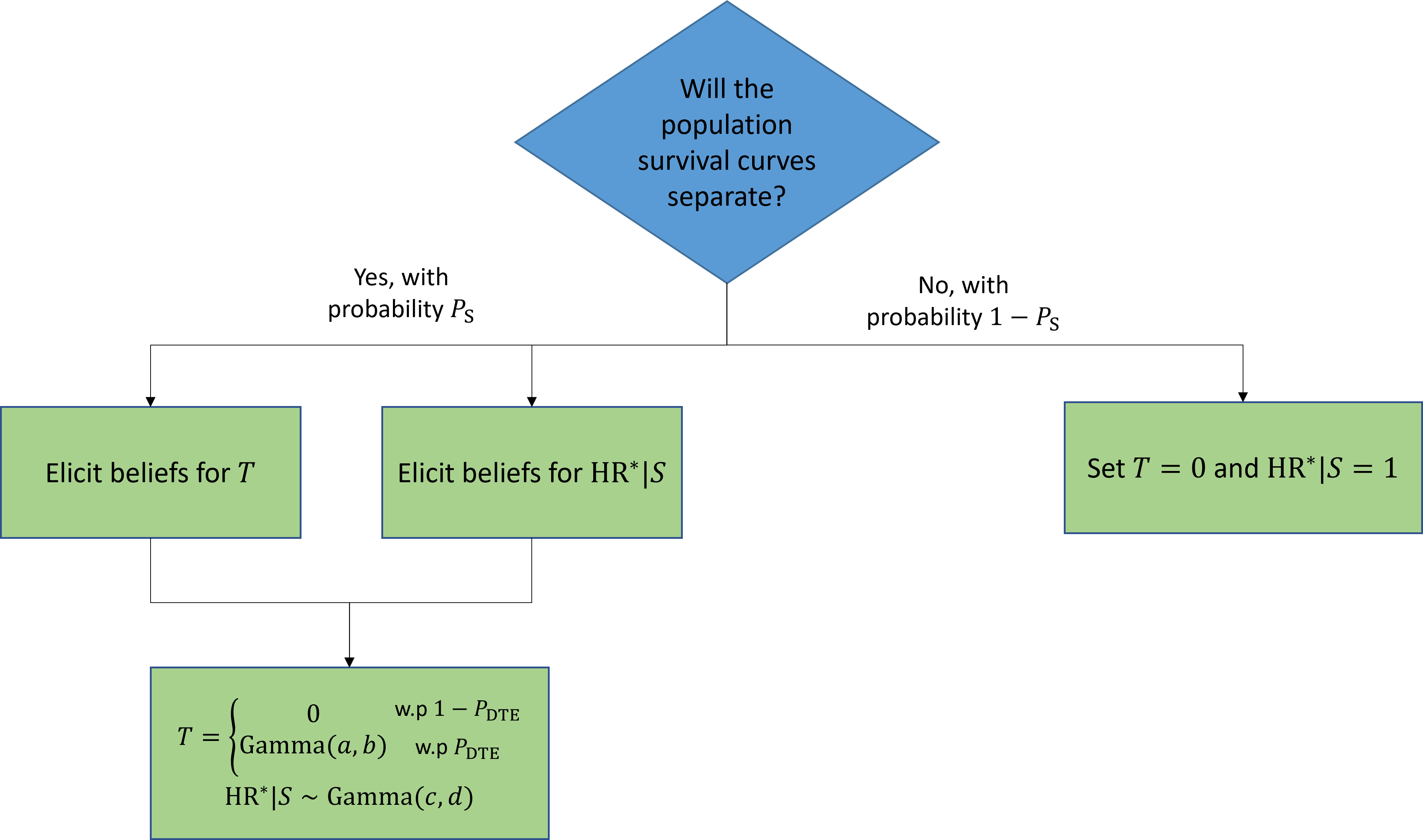}
    \caption{The proposed elicitation scheme as described in Sections \ref{sec:priorT} and \ref{sec:priorHR*}.}
    \label{fig:ElicitationProcedure}
\end{figure}

We propose a hierarchical procedure for eliciting judgements about $T$, $\lambda_e$ and $\gamma_e$, as shown in Figure \ref{fig:ElicitationProcedure}. The existence of a DTE presupposes that the treatment has any effect in the first place, but the experts may not be certain of this. Hence we first need to elicit a probability that the treatment has any effect, and then elicit judgements about $T$ conditional on the assumption that the treatment has some effect. To avoid ambiguity, we define ``a treatment effect'' as any separation between the survival curves for the control and treatment groups: they are not equal.  Hence the first question we ask is

\begin{quote}
\textit{``What is your probability that the population survival curves separate at some point in time?''}  
\end{quote}
We define $S$ to be the proposition that the population survival curves separate, and $P_S$ to be elicited probability that this proposition is true. Note that the proposition $S$ refers to the unobserved true population distributions of survival, and not sampled Kaplan-Meier curves that would be observed in a trial. We now elicit judgements about $T$, conditional on $S$. Given $S$, we allow for the possibility of no delay in the treatment effect; we propose a prior of the form 
\begin{equation}\label{eq:Tprior}
T = 
    \begin{cases}
    0, & \text{with probability } P_\text{DTE} \\
    D_\text{delay}, & \text{with probability } 1-P_\text{DTE}
\end{cases},
\end{equation}
with $D_\text{delay} \sim \text{Gamma}(a,b)$. Any non-negative distribution could be used for $D_\text{delay}$ but we expect the Gamma distribution to be sufficiently flexible. We therefore need questions that an expert would be willing to answer and from which we can identify values for $P_\text{DTE}$, $a$ and $b$. We elicit judgements about the probability $P_\text{DTE}$ by asking the following question:

\begin{quote}
\textit{``If we suppose that the population survival curves separate at some time, what is your probability that there is a delay before they separate?''}  
\end{quote}

Finally, we elicit judgements about the distribution of $D_\text{delay}$ by stating

\begin{quote}
\textit{``Suppose that the population survival curves separate with a delay. We want you to consider your uncertainty about the delay.''}
\end{quote}

We then use a standard method for eliciting a univariate distribution for $D_\text{delay}$. Methods for eliciting univariate distributions can be found in O'Hagan et al\cite{OHagan2006} and implemented using the Sheffield Elicitation Framework (SHELF).\cite{Shelf2010} SHELF is a package of protocols, templates and guidance documents for conducting expert elicitation. There are various methods that SHELF uses to elicit distributions that involve asking an expert to provide quantile judgements (e.g. a median; tertiles) or probability judgements (the probability of the uncertain quantity lying in some interval). In either case, the expert is, in effect, specifying points on their cumulative distribution function. Parametric distributions are fitted to these judgements using a least squares procedure: the parameters are chosen to ensure the points on the fitted cumulative distribution function are as close as possible to the elicited. Feedback (additional quantiles or probabilities from the fitted distribution) is then provided to the expert to check the adequacy of the elicited distribution.

\subsubsection{\texorpdfstring{Prior(s) for $\lambda_e$ and $\gamma_e$}{Prior(s) for lambda\_e and gamma\_e}}\label{sec:priorHR*}

The final two parameters which we need to elicit distributions for are the two treatment parameters, $\lambda_e$ and $\gamma_e$. We would not expect an expert to make judgements about these parameters directly. We instead follow usual practice of eliciting judgments about observable quantities,\cite{Kadane1998} from which a prior for $\lambda_e$ and $\gamma_e$ can be inferred. Some possible choices of observable quantities are

\begin{itemize}
    \item median survival time on the experimental treatment;
    \item survival probability at time $t$; and
    \item greatest distance between survival curves and how big is this difference.
\end{itemize}

In practice, we have found experts have a preference to make judgements about hazard ratios, for example,

\begin{itemize}
    \item hazard ratio at time $t$; and
    \item maximum hazard ratio and when this occurs.
\end{itemize}

To elicit $\lambda_e$ and $\gamma_e$, we require the expert to provide their beliefs for at least two of the above questions, which is likely to be a difficult task for the expert. We can simplify the elicitation task by making the assumption that $\gamma_e = \gamma_c$. We then have a piecewise-constant hazard ratio 
\begin{equation}\label{eq:HR2}
\text{HR}(t) = 
\begin{cases}
1, & t \leq T \\
\bigg(\frac{\lambda_e}{\lambda_c}\bigg)^{\gamma_c} , & t > T   
\end{cases}.
\end{equation}

\noindent We can rearrange Equation \ref{eq:HR2} for the case when $t>T$ to obtain
\begin{equation}\label{eq:lambdat}
\lambda_e = \lambda_c\text{HR}^{\frac{1}{\gamma_c}}.
\end{equation}
Hence, conditional on $\lambda_c$ and $\gamma_c$ we can elicit a distribution for the hazard ratio for $t>T$, from which a distribution for $\lambda_e$ can be derived. We make a standard modelling assumption that the treatment effect as described by the hazard ratio is independent of the control group response as determined by the parameters  $\lambda_c$ and $\gamma_c$. We investigate the implications of assumption $\gamma_e = \gamma_c$ in Section \ref{sec:simpgamma}.

We denote the post-delay hazard ratio by $\text{HR}^*$ (where it is assumed that $S$ is true). We propose a prior 
\begin{equation}\label{eq:HRprior}
\text{HR}^{*}|S \sim  \text{Gamma}(c,d).
\end{equation}

\noindent Again, any non-negative distribution could be used for $\text{HR}^{*}$. As with $T$, we need questions that an expert would be willing to answer and from which we can identify values for $c$ and $d$. We elicit judgements about the distribution $\text{HR}^{*}$ by stating

\begin{quote}
\textit{``Suppose that the population survival curves separate. We now want you to consider your uncertainty about the hazard ratio once the experimental treatment begins to take effect.''}
\end{quote}

\noindent We would then use a standard method for eliciting a univariate distribution for $\text{HR}^{*}$, as described in Section \ref{sec:priorT}.

Conditional on $S$ and any data $\boldsymbol{x}_{\text{hist}}$ related to control group survival, we assume a joint distribution of the form
\begin{equation}\label{eq:JD1}
\pi(T, \text{HR}^*, \lambda_c, \gamma_c|S, \boldsymbol{x}_{\text{hist}}) = \pi( \lambda_c, \gamma_c|S, \boldsymbol{x}_{\text{hist}})\pi(T|S)\pi(\text{HR}^*|S),
\end{equation}
with $P_S$ completing the prior specification. For algorithmic convenience, if $S$ is not true, we set $T=0$ and $\text{HR}^*=\text{HR}=1$. We do not expect $S$ to be informative for the control group survivor function, so we assume $\pi( \lambda_c, \gamma_c|S,\boldsymbol{x}_{\text{hist}}) = \pi( \lambda_c, \gamma_c|\boldsymbol{x}_{\text{hist}})$. We have assumed an expert's judgements about $\text{HR}^*|S$ are conditionally independent of $T$ given S. If the expert wanted to incorporate dependence between these parameters, a more complicated elicitation method could be used, such as the SHELF extension method,\cite{Shelf2010} illustrated in Holzhauer et al.\cite{Holzhauer2022}.

The elicitation technique discussed above assumes a single expert, but it is likely that in practice multiple experts will be consulted. Eliciting a distribution from multiple experts typically involves either eliciting a distribution from each expert separately and then aggregating the results, or alternatively getting the experts to agree on a single distribution. The SHELF method involves a combination of the two: experts first make judgements independently, which are then shared with the group. Following a facilitated discussion, the experts are then asked to agree on a single distribution reflecting the perspective of a ``Rational Impartial Observer''. Other methods for eliciting distributions from multiple experts are available.\cite{Cooke1991, Hemming2018} 

\subsection{Computing assurance under the DTE model}

We use these elicited distributions to calculate assurance for various sample sizes using Algorithm \ref{alg:ass}. This algorithm incorporates free parameters that can be adjusted to reflect operational constraints in a clinical trial: the control and treatment group sample sizes $n_c$ and $n_e$, and the total number of required events $E$. We let $E$ be a free parameter in the algorithm as it is common to run event driven survival trials. This is because, for a time-to-event endpoint, statistical information for the log-hazard ratio is a function of $E$ and therefore attained power will be determined by the number of events observed at the analysis time. Changing these free parameters will have consequences for assurance, so it is important to consider different combinations of trial designs in order to find the one which best suits the needs of the sponsor. For example, for a fixed $n_c$ and $n_e$, if we increase $E$ (the number of events) we may increase assurance at the cost of needing to run the trial for longer.

The recruitment schedule and analysis technique in Algorithm \ref{alg:ass} (and Algorithm \ref{alg:flexass} in Section \ref{sec:flexass}) are left unspecified, as these choices are not part of the assurance methodology; they can be selected separately. In the example of Section \ref{sec:example}, we use a Fleming-Harrington weighted log-rank test for the analysis (as there is high prior belief that the separation of the survival curves will be subject to a delay). By default, we assume uniform recruitment for 12 months.

To implement our methods we have developed an R Shiny app which is available both as an offline R package and hosted online. Instructions are provided in the Appendix. The Shiny app allows users to choose from two  recruitment schedules: piecewise constant and power method (taken directly from the \texttt{nphRCT}\cite{nphRCT2022} R package).
Different testing approaches have been proposed for use in the non-proportional hazards setting (e.g. max-combo test,\cite{Mukhopadhyay2022} weighted log-rank tests,\cite{Fleming1981} difference in RMST,\cite{Royston2013} and more\cite{Horiguchi2020}). Our app offers two of these statistical tests: a standard log-rank test and a Fleming-Harrington weighted log-rank test taken from the \texttt{nph}\cite{Ristl2021} R package). Finally, the elicitation process may inform refinements to the analysis plan.

\begin{algorithm}
\caption{calculating assurance when a DTE is likely to be present in a clinical trial}\label{alg:ass}
Inputs: sample sizes $n_c$ and $n_e$, the elicited priors $\pi( \lambda_c, \gamma_c| \boldsymbol{x}_{\text{hist}})$, $\pi(T|S)$, $\pi(\text{HR}^*|S)$, the probability of the survival curves separating $P_S$, the number of events $E$ (we require $E \leq  n_c+n_e$) and the number of iterations $N$. \\
For $i=1,\ldots, N$:

\begin{enumerate}
    \item sample $\lambda_{c,i}$ and $\gamma_{c,i}$ from $\pi(\lambda_c, \gamma_c)$;
    \item set $\gamma_{e,i} = \gamma_{c, i}$;
    \item sample survival times for the control group $x_{1,i},\ldots,x_{n_c,i}$ using $\lambda_{c,i}$, $\gamma_{c,i}$ (can use \texttt{rweibull()});
    \item sample $u$ from \texttt{runif(0, 1)}
    \item if $u < P_S$:
    \begin{itemize}
        \item sample $T_i$ and $\text{HR}^*_i$ from $\pi(T)\pi(\text{HR}^*|S)$;
    \end{itemize}
    else: \begin{itemize}
        \item set $T_i$ = 0 and $\text{HR}^*_i$ = 1;
    \end{itemize}
    \item transform $\text{HR}^*$ to $\lambda_{e,i}$ using Equation \ref{eq:lambdat};
    \item sample survival times for the experimental treatment group $y_{1,i},\ldots,y_{n_e,i}$ using $T_i$, $\lambda_{e,i}$ and $\gamma_{e,i}$ (can use inversion sampling via Equation \ref{eq:treatment});
    \item sample recruitment times $R_{1,i},\ldots,R_{n_c+n_e,i}$ from the pre-specified recruitment schedule;
    \item add the survival times from each group to the recruitment times to obtain a pseudo event time $P_{1,i},\ldots,P_{n_c+n_e,i}$;
    \item order the pseudo event times and define $E_T$ to be the time at which $E$ events have been observed;
    \item remove any observation in which the recruitment time $R_{j,i}$ > $E_T$;
    \item censor any observation for which the pseudo event time $P_{j,i}$ > $E_T$;
    \item for any censored observation, redefine the survival time to be $E_T - R_{j,i}$;
    \item perform the method of analysis on the data $x_{1,i},\ldots,x_{n_c,i}$ and $y_{1,i},\ldots,y_{n_e,i}$;
    \item define $U_i = 1$ if the data give rise to a `successful' outcome (0 otherwise).
\end{enumerate}
The assurance is then estimated as
\begin{equation*}
    \hat{P}(R) = \frac{1}{N}\sum^N_{i=1}U_i.
\end{equation*}
\end{algorithm}

\section{Example}\label{sec:example}
In this section, we illustrate the proposed method with a hypothetical example where we design a two-arm Phase III superiority trial to test whether a new drug is beneficial versus the current standard of care, docetaxel, in patients with advanced non–small-cell lung cancer (NSCLC). As the drug is in the IO area, we expect a DTE. The primary efficacy endpoint is OS, we assume uniform recruitment for 12 months, 1:1 allocation, and that the data will be analysed with a Fleming-Harrington weighted log-rank test with $\rho = 0$ and $\gamma=1$, as we have a high probability that the treatment will be subject to a delay and we want to place more weight on late differences in the survival curves. We assume that the trial final analysis will take place when 80\% of patients have died.

\subsection{Prior distribution(s) for the control parameters}\label{sec:controlexample}
\begin{figure}
\centering
\includegraphics[width=0.7\textwidth]{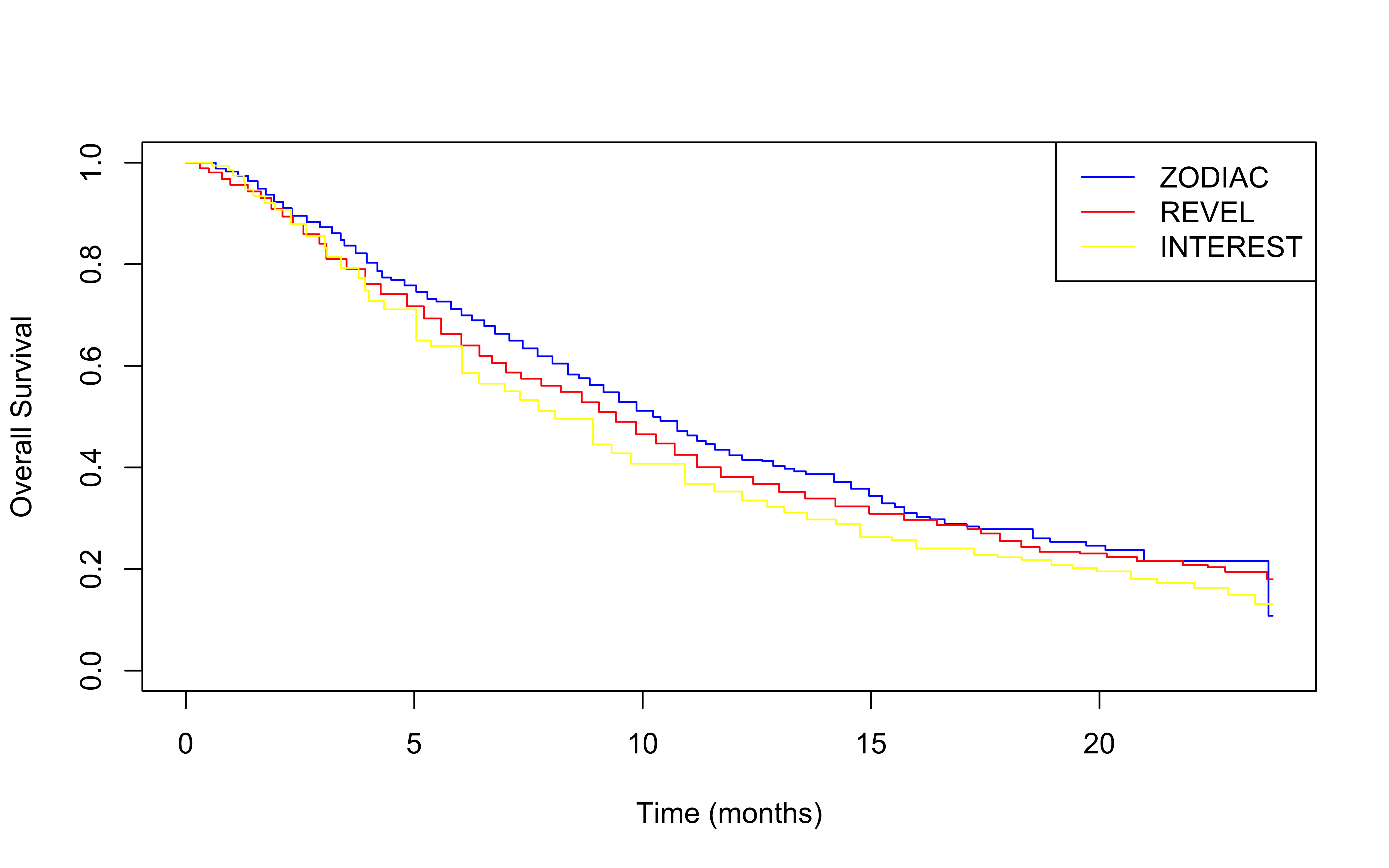}
\caption{Reconstructed Kaplan-Meier curves for the docetaxel arm in three different trials: ZODIAC,\cite{Herbst2010} REVEL\cite{Garon2014} and INTEREST.\cite{Kim2008} We see that the three curves are similar and we assume exchangeability of the trials in our example.}
\label{fig:ExampleCombinedControl}
\end{figure}

There exists historical data on docetaxel, so we are able to use this to generate a prior distribution for control group parameters. In Bertsche et al\cite{Bertsche2019} they found three trials in which docetaxel was used as the control in a clinical trial; ZODIAC,\cite{Herbst2010} REVEL\cite{Garon2014} and INTEREST.\cite{Kim2008} We also use the results from these three trials, but we use the published Kaplan-Meier curves to reconstruct the individual patient data.\cite{Liu2021} The three Kaplan-Meier curves can be seen in Figure \ref{fig:ExampleCombinedControl}. Since survival in all three trials appears similar, we choose to pool the data from all three trials and use this to update non-informative priors for $\lambda_c$ and $\gamma_c$, using Markov chain Monte Carlo (MCMC) to sample from the posterior distributions. The generated MCMC samples are then used as a prior distribution for the future trial of interest. 

\subsection{Eliciting the prior distribution for the length of delay}\label{sec:elicitT}

We now need to elicit the expert's probability that the population survival curves separate at some point in time. Suppose the expert specifies this as 90\%, that is $P_S$ = 0.9. The expert is then asked for their uncertainty about the length of delay, given that the survival curves do separate. Suppose the expert's probability that the effect of the experimental treatment will be subject to a delay is 70\%, that is, $P_\text{DTE}$ = 0.7. The expert is then for about their beliefs about $T$, conditional on there being a delay. The expert provides a median of 4 months and two quartiles (25\% and 75\%) of 3 and 5 months, respectively. A Gamma($a,b$) distribution is fitted to these judgements, so that $D_{\text{delay}}$ = Gamma(7.29, 1.76). Combining these beliefs using Equation \ref{eq:Tprior}, we have the following mixture prior distribution 
\begin{equation*}
T = 
    \begin{cases}
    0, & \text{with probability } 0.3 \\
    \text{Gamma(7.29, 1.76)}, & \text{with probability } 0.7
\end{cases}.
\end{equation*}

\noindent The fitted quartiles of a Gamma(7.29, 1.76) distribution are 3.03, 3.95 and 5.05. These would be presented to the expert for feedback.

\subsection{Eliciting the prior distribution for the post-delay hazard ratio}
The second quantity of interest is $\text{HR}^*|S$. Suppose the expert provides a median of 0.6 and two quartiles (25\% and 75\%) of 0.55 and 0.7, respectively. Again, we fit a Gamma($c,d$) to these judgements, so $D_\text{HR}$ = Gamma(29.6, 47.8). As per Equation \ref{eq:HRprior}, we have the following prior distribution
\begin{equation*}
\text{HR}^*|S \sim \text{Gamma(29.6, 47.8)}.
\end{equation*}

\noindent The fitted quartiles of Gamma(29.6, 47.8) distribution are 0.54, 0.61 and 0.69, and again, this would be presented to the expert for feedback.

\subsection{Calculating assurance}\label{sec:estass}
\begin{figure}
\centering
\includegraphics[width=0.7\textwidth]{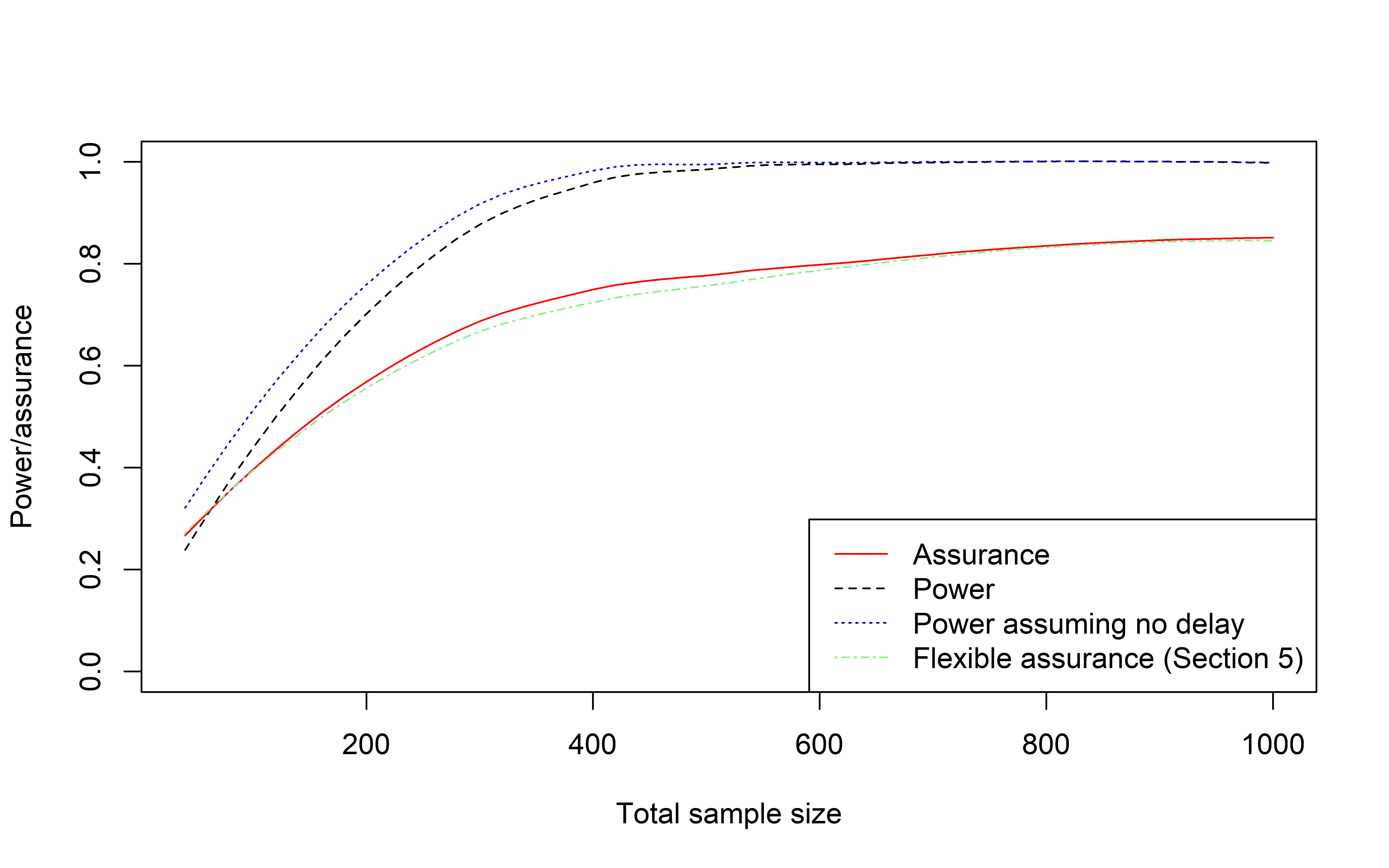}
\caption{Power/assurance curves for the example given in Section \ref{sec:example}. We see that the sample size required for 80\% power/assurance is greatly different under the different scenarios and highlights the importance of including  uncertainty in the design stage of a clinical trial}
\label{fig:PowerAss}
\end{figure}

We use these elicited prior distributions to calculate assurance for this example using Algorithm \ref{alg:ass}. In Figure \ref{fig:PowerAss}, an assurance curve is plotted to inform sample sizes required for this clinical trial. Also seen in Figure \ref{fig:PowerAss} are three other power/assurance curves. The two power curves correspond to including no uncertainty in the parameters, with the control parameters, $\lambda_c$ and $\gamma_c$, being the MLE from the three pooled data sets, as discussed in Section \ref{sec:controlexample}. The values for $T$ and $\text{HR}^*$ are the median values given by the experts. For one of the power curves, we have assumed that $T$ is 0, and therefore does not account for the fact that the treatment effect may be subject to a delay. Also shown is an assurance curve, corresponding to a more flexible approach to calculating assurance, this approach is presented in Section \ref{sec:flexass}. The distributions/values for the first three curves are found in Table \ref{tab:powerass}. We kept the recruitment scheduling, analysis method etc, as described at the start of Section \ref{sec:example}, constant across all four scenarios.

In Figure \ref{fig:PowerAss}, we see that both of the power calculations are much more optimistic than the other two scenarios; we require far fewer patients for the same power, at all sample sizes. This highlights the importance of incorporating uncertainty into the trial parameters. However, we must reiterate, the assurance method is not simply used for setting sample sizes for the proposed trial. We anticipate the assurance method being used as one step in a thorough process to decide whether or not to go ahead with the trial, and if we do run the trial, define the characteristics of the proposed trial; length, number of patients, number of events etc. For example, if we required a quicker trial, we may choose to decrease the number of events we need to observe before stopping the trial, $E$, but this will surely come at a cost of reducing the assurance/power seen. Therefore, it is important that a number of different trial designs are considered, and then assurance curves can be plotted to help inform the ultimate decision(s). 

\begin{table}[h]
\centering
\begin{tabular}{ccccccc}
\toprule
Calculation & $\lambda_c$ & $\gamma_c$ & $T$ & $\text{HR}^*$ & $P_S$ & $P_\text{DTE}$
\\ 
\midrule
Assurance & MCMC sample & MCMC sample & 
    $\text{Ga}(7.29, 1.76)$ & $\text{Ga}(29.6, 47.8)$ & 0.9 & 0.7
 \\ 
Power &  0.074 & 1.21 & 4 & 0.6 & 1 & 1 \\
Power assuming no delay & 0.074 & 1.21 & 0 & 0.6 & 1 & 0 \\ 
\bottomrule
\end{tabular}
\caption{Distribution/values for the parameters, for the first three scenarios seen in Figure \ref{fig:PowerAss}.}
\label{tab:powerass}
\end{table}

\section{Simplified prior distribution: discussion}\label{sec:simpgamma}

To elicit the parameters in this scenario, we select a method that would be both effective and straightforward for the experts. As a result, we simplified the original parameterisation by fixing $\gamma_e = \gamma_c$. By doing so, we were able to focus on hazard ratios as the basis for our questioning, thus ensuring that the elicitation process was both easy and intuitive for the experts.

However, it's important to investigate the robustness of this simplification. The results of our investigation are presented in the following section. Finally, we provide an alternative method for calculating assurance. This alternative is designed to accommodate situations in which the aforementioned simplification may not be preferred.

\subsection{Robustness of the parameterisation}

\begin{figure}[h]
    \centering
    \includegraphics[width=1\textwidth]{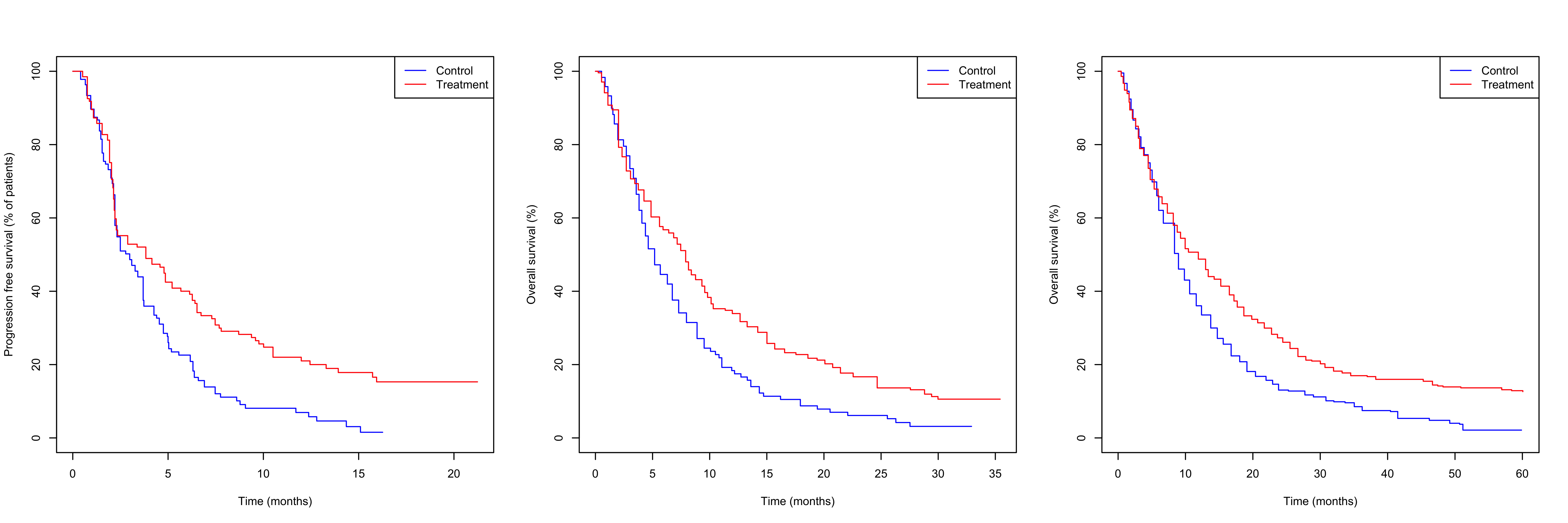}
    \caption{Kaplan-Meier plots for the data sets introduced in Section \ref{sec:simpgamma}. For (a) the data set is from trial Checkmate 017,\cite{Brahmer2015} for (b) the data set is from trial Checkmate 141\cite{Yen2020} and for (c) the data set is from trials Checkmate 017 and Checkmate 057 combined.\cite{Borghaei2021}}
    \label{fig:3KMPlots}
\end{figure}

The investigation aimed to assess the impact of the simplification we introduced, $\gamma_e = \gamma_c$, into the model by comparing two parameterisation methods: Method A and Method B. Method A incorporated the simplification, while Method B allowed $\gamma_e$ to vary. Historical data from three clinical trials (Checkmate 017,\cite{Brahmer2015} Checkmate 141,\cite{Yen2020} and Checkmate 017 and Checkmate 057 combined,\cite{Borghaei2021} the Kaplan-Meier plots for these trials are seen in Figure \ref{fig:3KMPlots}) with observed DTE were used to estimate the five unknown parameters (from Equations \ref{eq:control} and \ref{eq:treatment}) using both methods. For clarity, Table \ref{tab:FlexMethods} shows how the two methods estimate the five unknown parameters. 

\begin{figure}[h]
    \centering
    \includegraphics[width=1\textwidth]{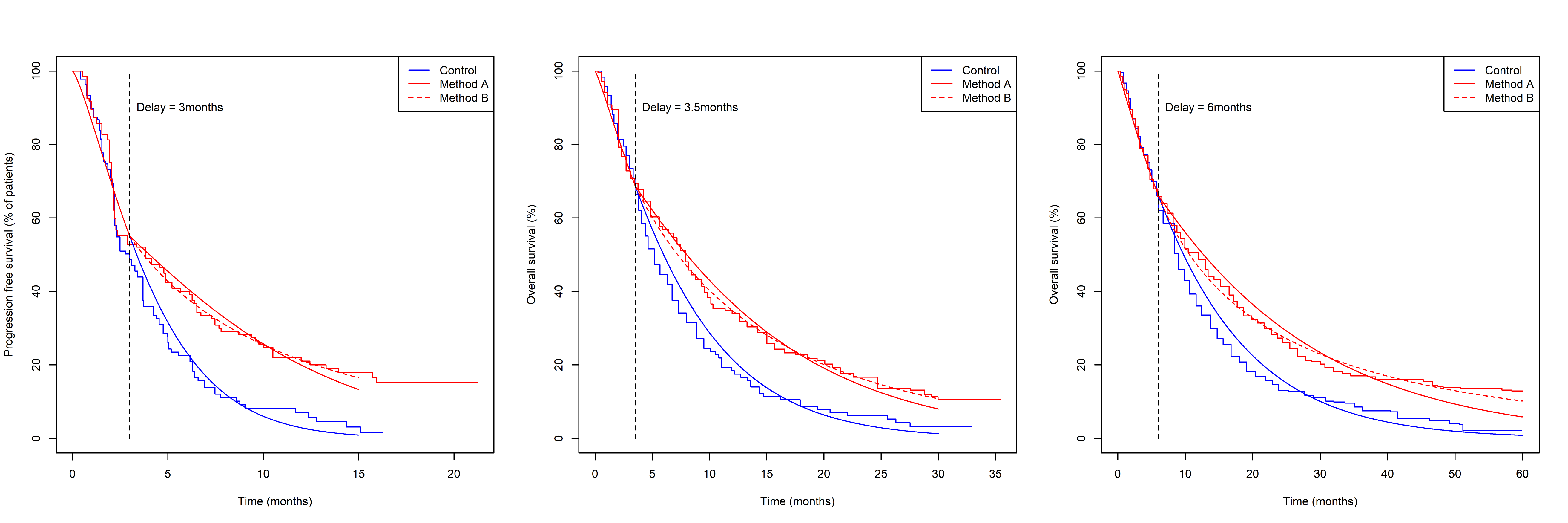}
    \caption{The same Kaplan-Meier plots as in Figure \ref{fig:3KMPlots}, with estimates for both Method A and B (as introduced in Section \ref{sec:simpgamma}) overlaid. For each data set, $T$ has been visually estimated. The control parameters, $\lambda_c$ and $\gamma_c$, have been estimated from the data (the overlaid blue line). In Method A, the simplification has been made ($\gamma_e = \gamma_c)$ and then $\lambda_e$ has been estimated using a least squares procedure. In Method B, both the treatment parameters ($\lambda_e, \gamma_e)$ have been simultaneously estimated.}
    \label{fig:KMAll3Estimates}
\end{figure}

The estimated parametric survival curves generated by both methods are presented in Figure \ref{fig:KMAll3Estimates}. Observing all three examples, we can see that the parametric treatment survival curve produced by Method B exhibits a marginally superior fit compared to the treatment curves derived from Method A. This is what we would intuitively expect, as Method B incorporates two free parameters, $\gamma_e$ and $\gamma_c$, while Method A only employs a single free parameter, $\gamma_c$. However, despite Method B giving a better fit to the data, Method A still approximates the data well. These findings suggest that the simplification introduced by Method A would likely have minimal practical impact on real decision-making processes.

\begin{figure}
\centering
\includegraphics[width=1\textwidth]{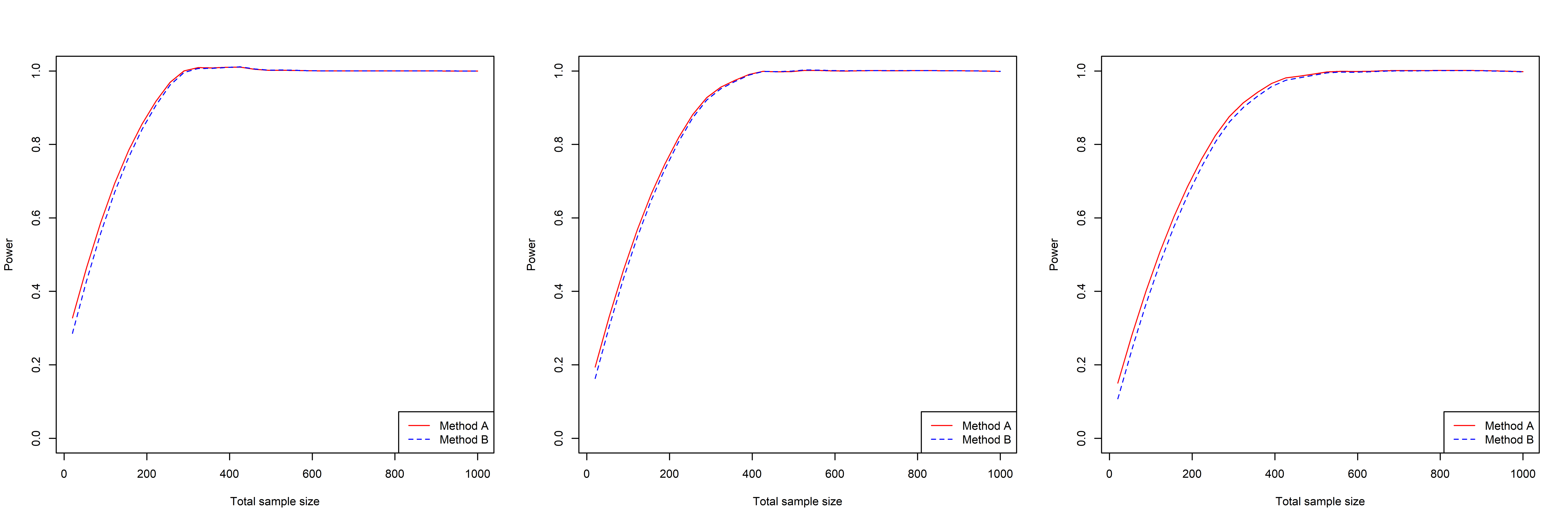}
\caption{Power curves for the two methods considered in Section \ref{sec:simpgamma}, for all three of the trials. We see that, in all three cases, the power curves are very similar to each other, thus indicating that the simplification does not seem make any practical difference in these examples. Method A is the simplification we made in the elicitation process.}
\label{fig:KMPower}
\end{figure}

Power calculations were performed to quantify the impact of the difference in the fitted experimental treatment survival curves. The results, depicted in Figure \ref{fig:KMPower}, showed almost indistinguishable power curves for both methods across all three datasets. This suggests that, in these examples, the assumption $\gamma_e = \gamma_c$ did not lead to different practical outcomes.

\begin{table}[h]
\centering
\begin{tabular}{ccccccl}
\toprule
\textbf{Method} & $\boldsymbol T$ & $\boldsymbol{\gamma_c}$ & $\boldsymbol{\lambda_c}$ & $\boldsymbol\gamma_e$ & $\boldsymbol\lambda_e$  \\ 
\midrule
A        & \multirow{2}{*}{Visually} & \multicolumn{2}{c}{\multirow{2}{*}{\texttt{survreg(dist = "weibull")}}} & $\gamma_c$ & MLE \\ 
B & & \multicolumn{2}{c}{} & MLE & MLE \\
\bottomrule
\end{tabular}
\caption{How each of the five parameters are estimated in both of the methods introduced in Section \ref{sec:simpgamma}, Method A is the simplification we made in the elicitation process (MLE = Maximum Likelihood Estimation).}
\label{tab:FlexMethods}
\end{table}

\subsection{A more flexible approach to evaluating assurance}\label{sec:flexass}
\begin{figure}[h]
    \centering
    \includegraphics[width=0.7\textwidth]{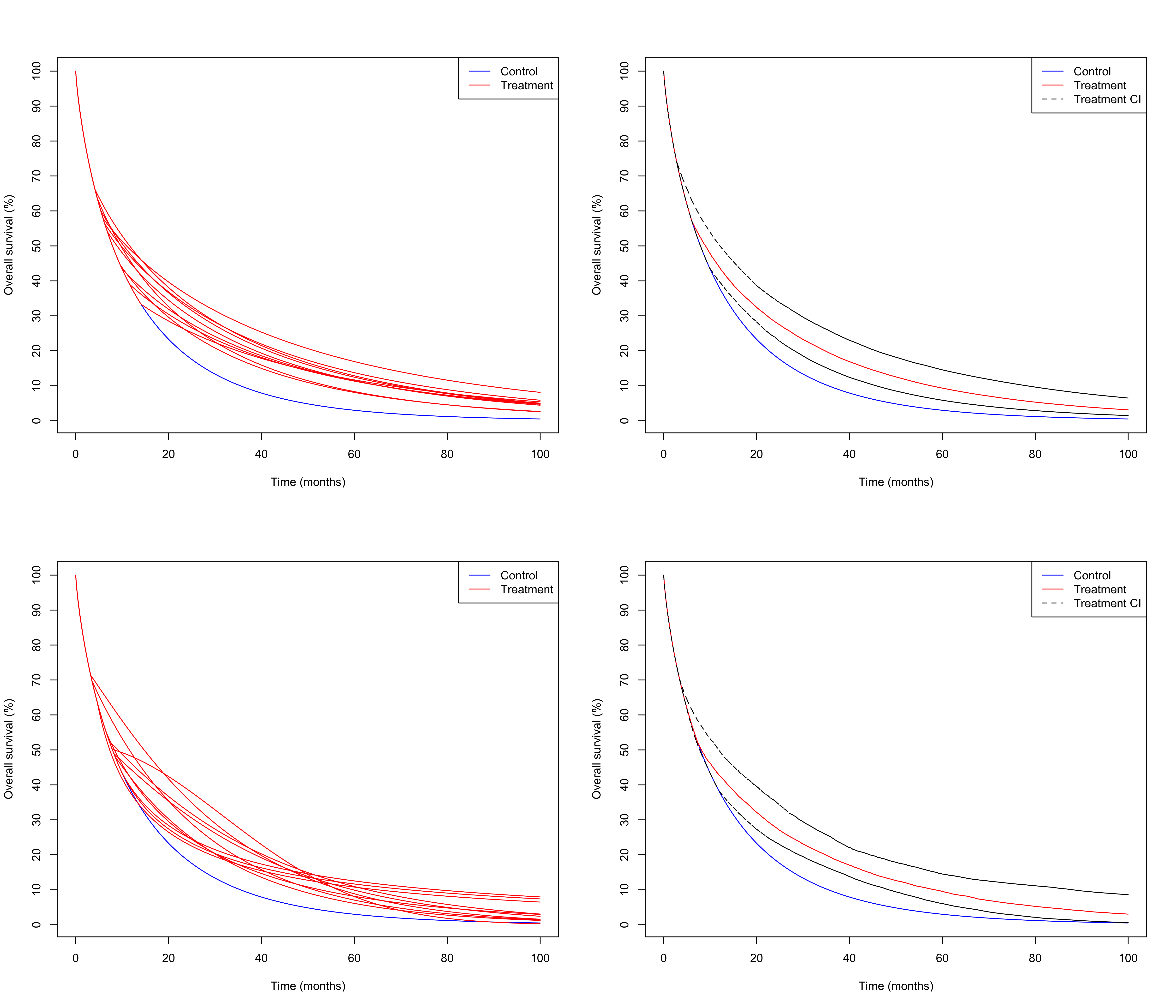}
    \caption{(a) 10 sampled experimental treatment survival curves, generated by Algorithm \ref{alg:ass}, (b) pointwise confidence intervals (0.1 and 0.9) for 500 of these sampled experimental treatment curves, (c) 10 sampled experimental treatment survival curves, generated by Algorithm \ref{alg:flexass}, (d) pointwise confidence intervals (0.1 and 0.9) for 500 of these sampled experimental treatment curves.}
    \label{fig:survCurves}
\end{figure}

We have demonstrated that for the three historical trials considered, the simplification does not appear to have any practical implications. However, by making this simplification, the possible experimental treatment survival curves are constrained to align with the shape of the control survival curve. In Figure \ref{fig:survCurves}(a), it can be observed that the experimental treatment survival curves seem to be `parallel' to the control curve due to this simplification and the fixed shape parameters, $\gamma_e = \gamma_c$, being the same for both curves.

\begin{figure}[h]
    \centering
    \includegraphics[width=0.7\textwidth]{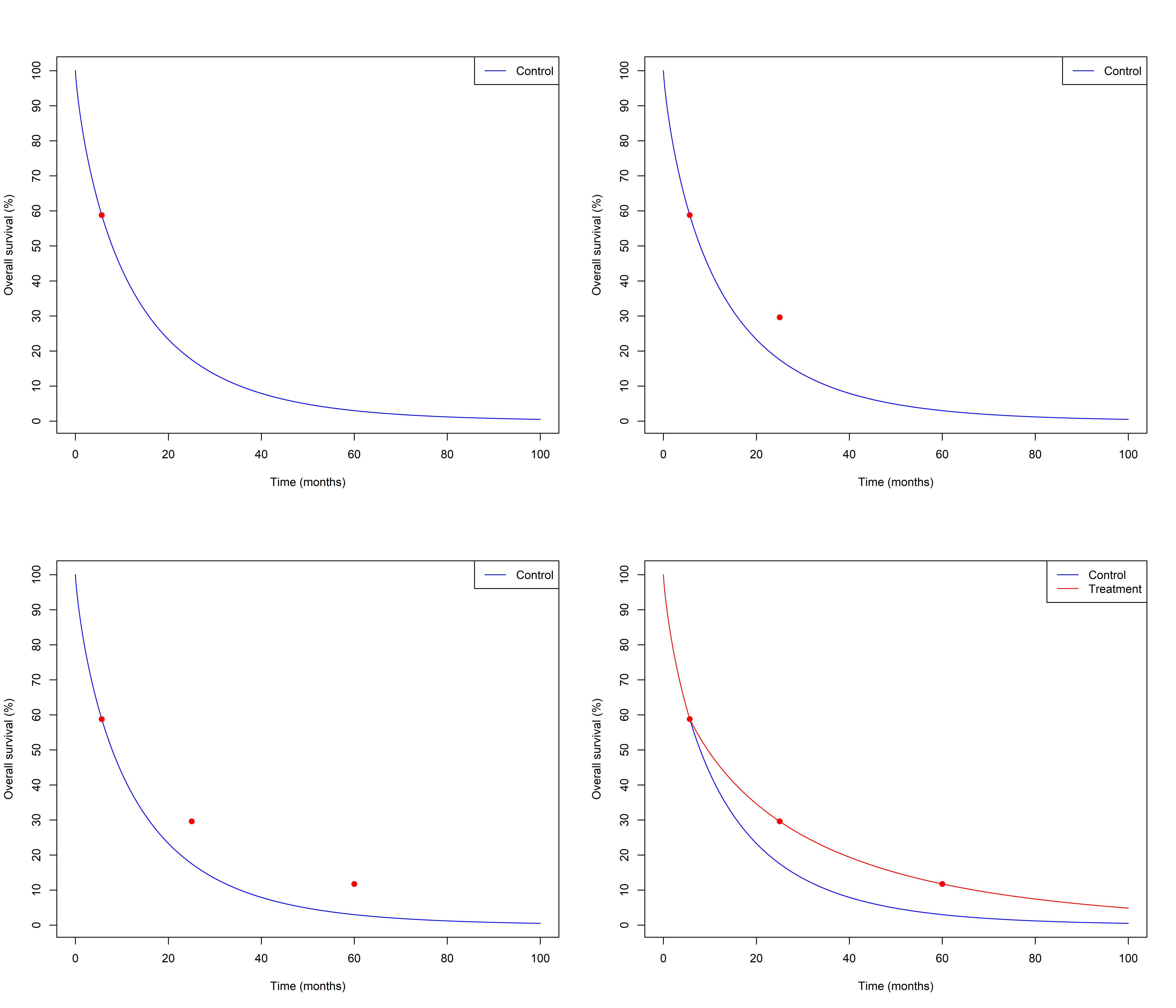}
    \caption{Showing the process of how an experimental treatment survival line is drawn for the flexible assurance. (a) a delay time is drawn from the prior distribution for $T$, (b) $s_1$ is sampled from all the survival probabilities at $0.25F$ through the trial, (c) $s_2$ is sampled from all the survival probabilities at $0.6F$ through the trial, (d) a piecewise Weibull distribution is fit to these sampled points.}
    \label{fig:FlexSeq}
\end{figure}

It is important to acknowledge that in certain trial designs, practitioners may feel uneasy about using this method, particularly if they believe that the experimental treatment survival curve would not align in a parallel manner to the control curve. In response to this concern, we have developed an alternative assurance calculation, referred to as Algorithm \ref{alg:flexass}. Algorithm \ref{alg:flexass} aims to generate experimental treatment curves that would be obtained if we had not made the simplification and allowed the curves to be sampled from a Weibull($\lambda_e, \gamma_e$) distribution after the delay occurred. The process is depicted in Figure \ref{fig:FlexSeq}.

To implement Algorithm \ref{alg:flexass}, we utilize the elicited mixture prior distributions for $T$ and $\text{HR}^*|S$. First, we sample a large number of experimental treatment curves, denoted as $M$. Next, we sample a value for $T$ from the elicited prior distribution, as shown in Figure \ref{fig:FlexSeq}(a). We define $F$ as the time at which the control curve reaches a survival probability of 0.01. Then, we independently sample two survival probabilities, $s_1$ and $s_2$, at 0.25$F$ and 0.6$F$ through the trial, respectively, as depicted in Figure \ref{fig:FlexSeq}(b) and \ref{fig:FlexSeq}(c). The only condition imposed is that $s_1>s_2$. Using $T$, $s_1$ and $s_2$, we apply a least squares procedure to fit a Weibull distribution to these points, as illustrated in Figure \ref{fig:FlexSeq}(d).

The sampled experimental treatment curves obtained from Algorithm \ref{alg:flexass} can be used to calculate assurance in the same manner as Algorithm \ref{alg:ass}. Figure \ref{fig:survCurves}(c) displays 10 sampled experimental treatment curves using this more flexible method. Additionally, Figure \ref{fig:survCurves}(d) shows the pointwise confidence intervals of the experimental treatment curves. Comparing these figures to Figures \ref{fig:survCurves}(a) and (b), we observe that the pointwise confidence intervals are very similar, indicating that the sampled experimental treatment curves fall within the same boundaries. However, the alternative method allows for sampling a more diverse range of curves, providing increased flexibility.

We have implemented Algorithm \ref{alg:flexass} in the example presented in Section \ref{sec:example}, and the results are seen in Figure \ref{tab:powerass}. The flexible assurance curve closely resembles the assurance curve obtained using Algorithm \ref{alg:ass}. This demonstrates that the more flexible assurance method may not significantly impact decision-making, but it may make practitioners more comfortable if they believe that the imposed constraint is not representative of the potential experimental treatment curves observed in practice. It is worth noting that the selection of time points (0.25$F$ and 0.6$F$) for sampling survival probabilities in Algorithm \ref{alg:flexass} is somewhat arbitrary. These values were chosen to produce realistic experimental treatment survival curves. However, if more restrictive or flexible curves are desired, these two points could be adjusted accordingly (e.g., closer together or further apart).

\begin{algorithm}
\caption{calculating flexible assurance for when a DTE is likely to be present}\label{alg:flexass}
Inputs: sample sizes $n_c$ and $n_e$,  the elicited priors $\pi( \lambda_c, \gamma_c| \boldsymbol{x}_{\text{hist}})$, $\pi(T|S)$, $\pi(\text{HR}^*|S)$, the probability of the survival curves separating $P_S$, the number of events $E$ (we require $E \leq  n_c+n_e$), the maximum trial length $L_{\text{max}}$, the number of initial samples $M$ and the number of iterations $N$. \\
\begin{enumerate}
    \item Initialise an empty matrix $A \in \mathbb{R}^{M\times t}$, where $t$ is the length of a vector \texttt{time = seq(0, $L_{\text{max}}$, by = 0.01)};
    \item for $j = 1, \ldots, M$:
    \begin{enumerate}[i]
       \item sample $\lambda_{c,j}, \gamma_{c,j}$ from $\pi(\lambda_c, \gamma_c)$; 
       \item set $\gamma_{e,j} = \gamma_{c, j}$;
       \item sample $u$ from \texttt{runif(0, 1)};
       \item if $u < P_S$:
       \begin{itemize}
           \item sample $T_j$, $\text{HR}^*_j$ from $\pi(T, \text{HR}^*|S)$;
       \end{itemize}
       else:
       \begin{itemize}
           \item set $T_j$ = 0,  $\text{HR}^*_j$ = 1;
       \end{itemize}
       \item transform $\text{HR}^*_j$ to $\lambda_{e,j}$ using Equation \ref{eq:lambdat};
       \item use $T_j$, $\lambda_{e,j}$, $\gamma_{e,j}$, and the control parameters $\lambda_c$, $\gamma_c$ to calculate the survival probability at each value of the \texttt{time} vector, using Equation \ref{eq:treatment};
       \item fill the $j$'th row of the matrix $A$ with these survival probabilities. 
    \end{enumerate}
    \item For $i=1,\ldots, N$:
    \begin{enumerate}[i]
    \item sample $\lambda_{c,i}, \gamma_{c,i}$ from $\pi(\lambda_c, \gamma_c)$;
    \item define $F$ to be the time at which the survival probability in the control group equals 0.01 (using $\lambda_{c,i}, \gamma_{c,i}$ and Equation \ref{eq:control});
    \item sample a survival probability, $s_{1,i}$, from the column in matrix $A$ which corresponds to $0.25F$;
    \item sample a survival probability, $s_{2,i}$ from the column in matrix $A$ which corresponds to $0.6F$ (we require $s_{2,i}<s_{1,i}$); 
    \item sample $T_i$ from $\pi(T)$;
    \item simultaneously solve these equations to find the best fitting values of $\lambda_{e,i}$ and $\gamma_{e,i}$ (can use \texttt{nleqslv()});
    \item sample survival times for the control group $x_{1,i},\ldots,x_{n_c,i}$ using the sampled $\lambda_{c,i}$, $\gamma_{c,i}$ (can use \texttt{rweibull()});
    \item sample survival times for the experimental treatment group $y_{1,i},\ldots,y_{n_e,i}$ using the sampled $T_i$, and $\lambda_{e,i}$, $\gamma_{e,i}$ (can use inversion sampling);
    \item sample recruitment times $R_{1,i},\ldots,R_{n_c+n_e,i}$ from from the pre-specified recruitment schedule;
    \item add the survival times from each group to the recruitment times to obtain a pseudo event time $P_{1,i},\ldots,P_{n_c+n_e,i}$;
    \item order the pseudo event times and define $E_T$ to be the time at which $E$ events have been observed;
    \item remove any observation in which the recruitment time $R_{j,i}$ > $E_T$;
    \item censor any observation for which the pseudo event time $P_{j,i}$ > $E_T$;
    \item for any censored observation, redefine the survival time to be $E_T - R_{j,i}$;
    \item perform the method of analysis on the data $x_{1,i},\ldots,x_{n_c,i}$ and $y_{1,i},\ldots,y_{n_e,i}$;
    \item define $U_i = 1$ if the data give rise to a `successful' outcome (0 otherwise).
    \end{enumerate}
    \end{enumerate}
    The assurance is then estimated as
    \begin{equation*}
    \hat{P}(R) = \frac{1}{N}\sum^N_{i=1}U_i.
\end{equation*}
\end{algorithm}

\section{Summary}\label{sec:summary}

In conclusion, assurance calculations have emerged as a valuable tool in the design and analysis of clinical trials. By incorporating Bayesian principles and considering prior distributions for unknown parameters, assurance calculations provide a more realistic assessment of a trial's probability of success compared to traditional power calculations. This approach acknowledges the inherent uncertainties in clinical research and allows for the simulation of trial outcomes based on sampled prior distributions. Assurance calculations offer several advantages for trial design and decision-making. They assist in optimizing sample size, assessing risks, and evaluating the effectiveness of different trial setups, including the timing and number of planned interim analyses. Furthermore, assurance evaluations enable better-informed go/no-go decisions regarding study conduct, directing resources towards research programs with the highest expected impact for patients.

In the rapidly evolving field of immuno-oncology, assurance calculations have the potential to address challenges associated with time-varying or delayed treatment effects on time-to-event endpoints. We have extended the assurance method to include survival trials in which a delayed treatment effect is likely to occur. Overall, assurance calculations provide a robust framework for quantifying the probability of success in clinical trials while considering uncertainty. By incorporating Bayesian methods and accommodating complexities in trial design, assurance calculations contribute to more informed decision-making, improved trial design, and ultimately, more effective and impactful clinical research.

\subsection*{ACKNOWLEDGEMENTS}
This work has been supported by a University of Sheffield EPSRC Doctoral Training Partnership (DTP) Case Conversion with Novartis Scholarship [project reference 2610753]. 

\subsection*{DATA AVAILABILITY STATEMENT}
Data sharing is not applicable to this article, as no new data were created or analysed in this study.

\appendix\label{sec:appendix}
An R\cite{R2022} package, \texttt{DTEAssurance}, for implementing the methods described in this paper is available on GitHub, at \texttt{https://github.com/jamesalsbury/DTEAssurance}. The website also includes an illustration of using the package to replicate the examples in this paper. This package is installed with the commands
\\ \\
\texttt{install.packages("devtools")}

\noindent \texttt{devtools::install\char`_github("jamesalsbury/DTEAssurance")}.
\\ \\
An app for implementing these methods, produced with \texttt{shiny},\cite{Shiny2022} can be used online at \texttt{ https://jamesalsbury.shinyapps.io/DTEAssurance/}. 
A version of the app for offline use is included in the \texttt{DTEAssurance} package.

\end{document}